\documentstyle[twocolumn,aps]{revtex}

\input{epsf}
\begin{document}

\draft
\title{Coherent control of spin squeezing}
\author{C. K. Law, H. T. Ng, and P. T. Leung}
\address{{Department of Physics,
The Chinese University of Hong Kong,}\\
{Shatin, NT, Hong Kong}}
\date{\today}
\maketitle
\begin{abstract}
We report an interaction that controls spin squeezing in  
a collection of spin $1/2$ particles. We describe how spin 
squeezing can be generated and maintained in time. Our scheme 
can be applied to control the spin squeezing in a
Bose condensate with two internal spin states.
\end{abstract}


Squeezed spin states (SSS) of atoms have reduced quantum fluctuations
that are useful in enhancing sensitivity in precision spectroscopy  
\cite{wineland}. Since the early work by Kitagawa and Ueda \cite{ueda}, 
there have been several proposals of generating SSS in different
configurations \cite{wineland,puri,polzik,molmer,xxx,bigelow}. In the original 
proposal, Kitagawa and  Ueda \cite{ueda} have identified two fundamental 
types of nonlinear spin interactions that lead to spin squeezing:
\begin{eqnarray}
&& H_1 = 2 \kappa J_z^2,
\\
&& H_{2} = i\kappa (J_{+}^2-J_{-}^2).
\end{eqnarray}
Here $J$s' are collective angular momentum operators and $\kappa$ describes 
the interaction strength. The physical realization of these interactions is 
still a challenge, but Milburn {\it et al.} have pointed out that $H_1$ can 
naturally be found in a weakly interacting Bose condensates \cite{Walls}. More 
recently, Sorensen {\it et al.} \cite{xxx} and Raghavan {\it et al.}
\cite{bigelow} made use of $H_1$ to predict spin squeezing in a spinor 
Bose condensated gas.

In this report we study squeezing effects generated by the Hamiltonian 
$(\hbar =1)$,
\begin{equation}
H_{3} = 2\kappa J_z^2+\Omega J_x.
\end{equation}
This Hamiltonian generalizes $H_1$ by adding a linear interaction term 
$\Omega J_x$, where the interaction strength $\Omega$ (assumed positive) can be 
controlled by an external field.  We note that Milburn {\it et al.} \cite{Walls} 
first proposed the model $H_3$ to study the dynamics of a Bose condensate in a 
double-well potential. In their work, they have shown that the presence of 
$\Omega J_x$ can significantly affect the mean-field dynamics. In this paper, 
we focus on the issue of spin fluctuations. We shall show that $H_3$ can 
generate {\em strong squeezing in an extended period of time}.

To be definite, we consider a $J-$spin system that can be regarded as a 
collection of $2J$ spin $1/2$ particles. Following Kitagawa and  Ueda's 
criteria of spin squeezing, we introduce the squeezing parameter,
\begin{equation}
\xi_s ={{\sqrt 2\left\langle {\left( {\Delta J_\bot } \right)_{min }
} \right\rangle} \over {J^{1/2}}},
\end{equation}
where ${\left\langle {\left( {\Delta J_\bot } \right)_{min}
} \right\rangle }$ is the smallest uncertainty 
of an angular momentum component perpendicular to the mean angular momentum
${\left\langle {\bf J} \right\rangle }$. A state is said to be a squeezed 
spin state if $\xi_s < 1$.

Let us first examine the exact numerical solutions of the time-dependent 
Schr\"odinger equation governed by the Hamiltonian $H_3$. We consider that the 
system  starts from the lowest eigenvector of $J_x$, 
$\left| {J,m_x=-J} \right\rangle $.
Such an initial state is favorable in generating spin squeezing because of 
the twisting effect due to the nonlinear interaction $2\kappa J_z^2$ along 
the $z-$axis. In Fig. 1 we show the typical behavior of the squeezing parameter 
as a function of time. In the case of $\Omega =0$ (i.e., the $H_1$ model) shown 
in Fig. 1a,  we see that $\xi_s$ reaches a minimum after a characteristic time. 
However, such a squeezing can only be maintained in a certain time period. 
As the time increases, the system is less squeezed and eventually becomes 
unsqueezed. 

The key advantage of the model $H_3$ ($\Omega \ne 0$) is the maintenance
of squeezing in an extended period of time. This is clearly shown in 
Fig. 1b-d. We see that $\xi_s$ can be kept below unity in a much longer period 
of time than that for the $\Omega=0$ case. Indeed, for the choice of $\Omega$ 
used in Figs. 1b-d, the system exhibits squeezing most of the time. The 
interaction strength $\Omega$ affects the minimal $\xi_s$ that the system 
can reach. Fig. 1c represents a near optimal choice of $\Omega$ for $J=100$. 
A closer look at the minimal value of $\xi_s$ indicates that it 
is less than (i.e., more squeezing) that in Fig. 1a.

\begin{figure}
\centerline{
\epsfxsize=3.2in
\epsfbox{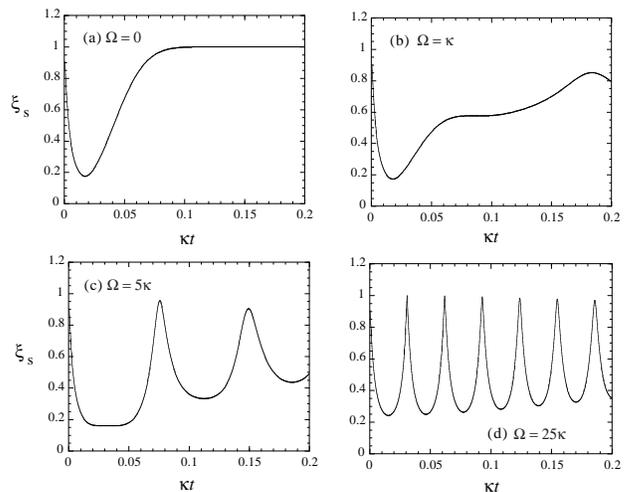}}
\vspace{5mm}
\caption{Typical time dependence of $\xi_s$ for the system starting from 
the initial state $\left| {J,m_x=-J} \right\rangle $. Here $J=100$ is used.}
\end{figure}

Another advantage of the interaction model $H_3$ is the maintenance of large 
coherent (mean) component of the collective spin. This feature is shown in 
Fig. 2 where the expectation values ${\left\langle {J_x} \right\rangle }$ are 
plotted against time for various values of $\Omega$. We remark that for the 
system starts from  $\left| {J,m_x=-J} \right\rangle$, the only nonvanishing 
spin component is $J_x$ because  
${\left\langle {J_y} \right\rangle }={\left\langle {J_y} \right\rangle }=0$
at all times. In Fig. 2, we see that for a sufficiently large $\Omega$ 
(for example the $\Omega=25 \kappa$ curve), 
${\left\langle {J_x} \right\rangle }$ changes slightly in the course of time. 
This is in contrast to the $\Omega=0$ case in which 
${\left\langle {J_x} \right\rangle }$ vanishes after some time. Since a strong 
coherent spin component is often needed in increasing the sensitivity of precision 
measurement (such as in Ramesy specroscopy \cite{wineland}), $H_3$ is more 
desirable in this regard. 

\begin{figure}
\centerline{
\epsfxsize=3.in
\epsfbox{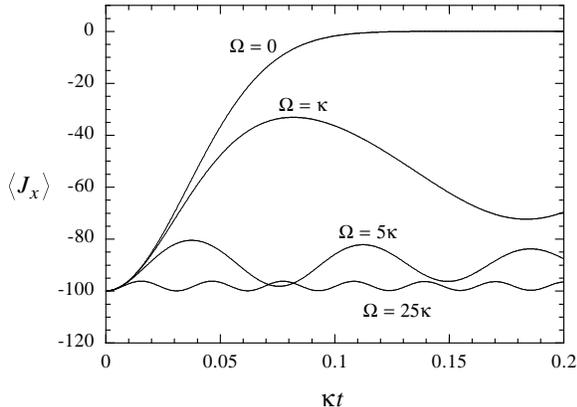}}
\vspace{5mm}
\caption{Time dependence of the expectation value of $J_x$, same parameters 
as in Fig. 1}
\end{figure}


Although exact analytic solutions of the nonlinear problem are not available, 
the squeezing behavior produced by $H_3$ can be understood when $\Omega$ is 
sufficiently larger than $\kappa$, i.e., $\Omega \gg \kappa$. 
First we recall that the system is prepared to start from 
the lowest eigenstate of $J_x$, $\left| {J,m_x=-J} \right\rangle $, i.e.,
\begin{equation} 
J_x  \left| {J,m_x=-J} \right\rangle = -J\left| {J,m_x=-J} \right\rangle .
\end{equation}
Such a state minimizes the energy assoicated with the interaction $\Omega J_x$. 
If $\Omega \gg \kappa $, the external field forces the total spin to remain 
polarized in the -ve $x-$direction because it costs energy to change the spin 
vector. This explains why a large coherent component of 
${\left\langle {J_x} \right\rangle }$ 
can be maintained. Now we look at the Heisenberg's equation of
motion of the angular momentum operators in the $y-$ and $z-$ directions,
\begin{eqnarray}
&& \dot J_z=\Omega J_y, \\
&& \dot J_y=-\Omega J_z+2\kappa \left( {J_zJ_x+J_xJ_z} \right).
\end{eqnarray}
Based on the fact that $J_x$ remains unchanged approximately, it is justified 
to make an approximation: Replacing $J_x$ by $-J$. We call such an approximation 
as a {\em frozen spin approximation} \cite{remark}. In this way, we have
\begin{equation}
\ddot J_z\approx -\left( {\Omega ^2+4\kappa \Omega J} \right)J_z ,
\end{equation}
which permits harmonic solutions,
\begin{eqnarray}
&& J_z(t)\approx J_z(0)\cos \omega t+\Omega J_y(0)\sin \omega t/\omega 
\\
&& J_y(t)\approx -\omega J_z(0)\sin \omega t/\Omega +J_y(0)\cos \omega t ,
\end{eqnarray}
where the frequency $\omega \equiv \sqrt {\Omega ^2+4\kappa \Omega J}$ is 
defined.

Eqs. (9) and (10) are operator solutions under the frozen spin approximation. 
The time-dependent spin fluctuations are given by,
\begin{eqnarray}
&& \left\langle {\left( {\Delta J_z(t)} \right)^2} \right\rangle 
 \approx \left\langle {J_z^2(0)} 
\right\rangle \cos ^2\omega t+{{\Omega ^2} \over {\omega ^2}}
\left\langle {J_y^2(0)} 
\right\rangle \sin ^2\omega t, \\
&& 
\left\langle {\left( {\Delta J_y(t)} \right)^2} \right\rangle \approx 
\left\langle {J_y^2(0)} \right\rangle \cos ^2\omega t+{{\omega ^2} \over 
{\Omega ^2}}\left\langle {J_z^2(0)} \right\rangle \sin ^2\omega t.
\end{eqnarray}
Here the cross terms $\left\langle {J_z(0)J_y(0)} \right\rangle$
and $\left\langle {J_y(0)J_z(0)} \right\rangle$  do not appear 
because they are identically zero with respect to the initial state (5).
Now using the fact that $\omega > \Omega$ and
\begin{equation}
\left\langle {J_z^2(0)} \right\rangle =\left\langle {J_y^2(0)} 
\right\rangle =J/2~,
\end{equation}
we find that reduced spin fluctuations occurs in the 
$z-$direction,
i.e.,
$\left\langle {\left( {\Delta J_z(t)} \right)^2} \right\rangle \le J/2$. 
In other words {\em the system is always squeezed} except at the times 
$t=n\pi / \omega$. 
The strongest squeezing occurs at $t=t^*=(2n+1)\pi /2\omega$ with
\begin{equation}
\left\langle {\left( {\Delta J_z(t)} \right)^2} \right\rangle _{t=t^*} 
\approx {{\Omega ^2 J} \over {2\omega ^2}}.
\end{equation}
This corresponds to the squeezing parameter at $t^*$,
\begin{equation}
\xi _{min } \equiv \left. {\xi _s} \right|_{t=t^*}
\approx {\Omega  \over \omega } < 1.
\end{equation}
From the definition of $\omega$ above, we see that the squeezing 
parameter $\xi_{min}$ is approximately ${(4\kappa J/\Omega)}^{-1/2}$ when 
$\kappa J \gg \Omega$. 
Therefore the system is less squeezed if $\Omega$ is large, but more squeezing 
can be achieved by increasing the number of particles. We should point out that 
the frozen spin approximation becomes less valid when $\Omega$ is comparable to 
$\kappa$. Nevertheless, the approximation captures the essential 
physical picture. We have compared the approximate analytical results with the 
exact numerical solutions, we found a good agreement in $\xi _{min }$ and
the oscillation frequency $\omega$ as long as $\Omega \gg \kappa$.

In order to to determine how optimal squeezing depends on $\Omega$ and 
particle number $2J$ beyond the frozen spin approximation, we have examined 
the exact numerical solutions for a wide range of $J$ and $\Omega$. 
In Fig. 3 we show the values of $\xi_{min}$ that can be attained by 
our model $H_3$ for different $J$s. These values of $\xi_{min}$ are attained 
by using optimal $\Omega$ for the corresponding $J$ (see the inset of Fig. 3). 
For example, we have found that for $J=500$, $\Omega \approx 10\kappa$ yields 
an optimal squeezing parameter $\xi_{min} \approx 0.09$. These numerical 
findings are consistent with the prediction from the frozen spin approximation 
that strong squeezing exhibits in the domain $\kappa J \gg \Omega  \gg \kappa$.

\begin{figure}
\centerline{
\epsfxsize=3.2in
\epsfbox{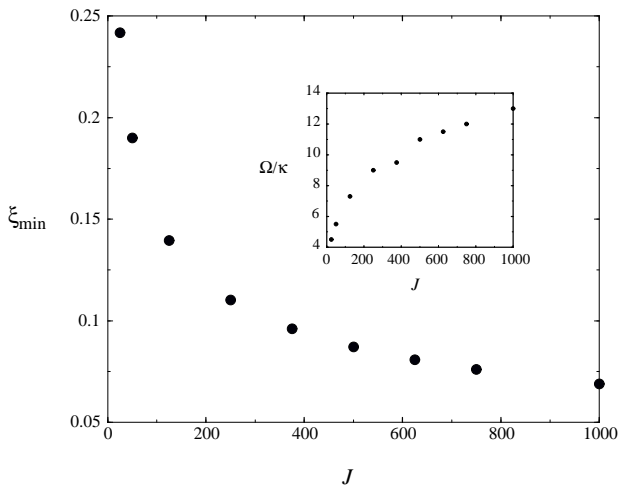}}
\vspace{5mm}
\caption{The minimum value of the squeezing parameter that can be attained 
for various $J$. The inset shows the optimal $\Omega$ used.}
\end{figure}

In conclusion, we have discovered how an external field can be applied to 
control spin fluctuations, which is expected to play an prominent role in the 
internal dynamics of Bose condensates. Our results indicate that a 
cooperation of the nonlinear self-interaction $2 \kappa J_z^2$ and the external 
interaction $\Omega J_x$ can generate spin squeezing in an extended period of time. 
In our scheme the coherent component of the collect spin can be locked in the 
$x-$direction, and the reduced fluctuations always appear in the $z-$direction. 
These advantages are not found in the standard model $H_1$. Finally we remark 
that spinor Bose condensates have an intrinsic $H_1$ type self-interaction among 
particles. The application of an external magnetic field or Raman fields can be 
used to prepare the required initial state and to realize the coupling 
$\Omega J_x$ \cite{xxx,bigelow}.

\acknowledgments 
C.K.L. thanks discussions with H. Pu, S. Raghavan, and N. P. Bigelow. 
We acknowledge the support from the Chinese University of Hong Kong Direct 
Grant (Grant Nos: 2060148 and 2060150). C.K.L. is supported by a postdoctoral 
fellowship at the Chinese University of Hong Kong.



\begin{references}
\bibitem{remark} Although the $x-$component of the spin is essentially frozen
by the external field, the $y-$ and $z-$ spin component can still evolve. 
It is the spin fluctuations on the $y-z$ plane that are needed to determine
the squeezing parameter.
\end{references}
\end{document}